# Carrier type modulation in current annealed graphene layers

**Pawan Kumar and Amit Kumar***

School of Material Science and Technology, Indian Institute of Technology (BHU), Varanasi - 221005, India

We report on the electrical properties of current annealed graphene and few layer graphene devices. It is observed that current annealing for several hours results the n-type doping in the graphene layers. After current annealing Dirac point start to shift toward positive gate voltage and saturate at some fixed gate voltage. N-type conduction in current annealed graphene layers is caused by the charge trapping in oxide layer during current annealing and recovery of charge neutrality point with time span is understood due to the de-trapping of charge with time.

*Corresponding electronic mail:  amitnsc@gmail.com

Remarkable band structure of graphene makes it a promising material for technological applications and fundamental physics interest.[1] High carrier mobility, ability to modulate carrier density in graphene bands and low cost device fabrication methods considered tremendous interest in investigating of its exceptional electronic properties and use in nanoelectronic applications.[2,3] In last few years, there has been much progress in the fabrication and understanding of graphene based devices (field effect transistor, atomic scale switches, high frequency devices etc).[4] Main hurdle has been to improve the carrier mobility by eliminating the charged impurities adsorbed on graphene device during fabrication process. Attempts have been made to improve graphene mobility by selecting different substrates[5] or annealing in inert atmosphere or vacuum in order to remove unintentional contamination/impurities.[6] Electrical current annealing is thought to be a better method to anneal graphene sample before measurement[7-10], where re-adsorption of atmospheric gases and water vapour can be removed. Recently, Scanning gate microscopy is used to explore the local conductivity in current annealed graphene flake.[11] It is reported that annealing at low current density exhibit micron size inhomogeneities whereas annealing at higher current density reduces the charge neutrality point inhomogeneities and increase inhomogeneities closed to electrodes.[11] However, the detail investigation of current annealing effects on graphene or oxide layer is still missing. To adopt current annealing method in improvement of the quality/mobility of graphene based devices a better understanding of current annealing process is required.

In this communication, we report a detailed investigation of the effects of graphene and few-layer graphene annealing on electrical transport. As expected, it is observed that annealing for several hours improve the sample's cleanliness, but also changes the overall system's characteristics so that a large negative electrostatic doping is required to recover charge neutrality point (CNP) of graphene. This effect is however reversible with time and points toward charging effects of the gate dielectric.

Graphene and few layer graphene samples are prepared by micro-mechanical exfoliation of natural graphite onto a $SiO_2/Si$ substrate with $SiO_2$ thickness of 300nm. The $SiO_2$ dielectric layer is used as a back-gate and allows a controlled electrostatic doping of the sample. Mono-layer or few layer graphene flakes are identified by means of optical microscopy, atomic force microscopy and micro-Raman scattering measurements. Electron beam lithography was used to deposit electrical contacts made of Ti (5nm)/Au (50nm) through a poly methyl methacrylate (PMMA) mask. Except the standard sample cleaning methods using alcohol, acetone and ionized water, the samples were not further processed by any additional solvents or chemicals. Two terminal AC electrical transport measurements were performed at room temperature in high vacuum ($<2 \times 10^{-6}$ Torr). The samples have been alternatively annealed using either a global heating treatment or making use of a large electrical current circulating through the device. The last process is referred to as electrical annealing.[7] In this case; a DC voltage is applied through the source-drain electrodes, at constant back-gate voltage, while the current flowing through the sample is monitored. The current injection is stopped when the current shows a noticeable change as compared to its initial value, indicating a resistance change of the sample. The resistance (R) versus back gate voltage (Vg) characteristics shown below has all been measured several hours after annealing, to allow the system to relax into equilibrium conditions at room temperature. Current annealing was performed on three graphene devices, called as sample A (mono-layer graphene), sample B (~4-5 graphene layers) and sample C (>6 graphene layers).

Figure 1-(a) shows the sample's resistance as a function of applied back-gate voltage (Vg) before and after successive annealing treatments for sample A. As fabricated, the sample is expected to be heavily contaminated with adsorbents, mainly from PMMA resist residues as well as water or oxygen due to direct exposure to air. The sample appears as heavily doped and the maximum resistance at the CNP is out of the back-gate range. Annealing at $100^0$C in vacuum for 20 hours certainly improved the sample's quality but did not remove all the adsorbed impurities, as suggested by highly p-doped R-(Vg) characteristics. On the other hand, successive current annealing results in significant improvements and the behaviour of pristine graphene is almost recovered. At intermediate stages, we observe two-peak R-(Vg) characteristics, which are assigned to the presence of local zones in-homogeneously doped (n-type and p-type) in graphene or to the presence of invasive contacts.[12-14] Non-equilibrium Green's function formalism based simulations suggest that the two peak structures may also be due to the non-constant density of states in the electrodes.[15] Finally, current annealing with current injection of 2.5 mA for 60 minutes results in a symmetric R-(Vg) curve centred at $V_{CNP}$ = -16.8 V suggesting n-type doping in graphene. As the sample is left in vacuum at room temperature, the CNP progressively shift toward positive gate voltage and is finally

stabilized at Vg=1V. Similar effects are observed in few layer graphene devices, namely in sample B and C as shown in figure 1 (b) - (c) respectively. However, the lower mobility and the reduced gating efficiency due to screening of the extra graphene layers prevent a detailed observation of the phenomena. Afterward we will mainly discuss results on monolayer graphene device (sample A).

Figure 2 show the sample's resistance as a function of applied back-gate voltage (Vg) for re-annealed sample A. The black curve [A] in figure 2 is the same as the last curve in figure 1 (a) with additional 6 hours of pumping. Re-current annealing leads to n-typing doping in graphene, CNP at -16.8 V (B-blue curve). It is noticed that just after current annealing the hysteresis is higher, which reduced with time and CNP move towards positive. It is also noticed that resistance at CNP increases with time suggest the reduction of residual/excess carrier concentration at CNP. These measurements show that n-type conductivity can be achieved by current annealing and reproducible in mono-layer and few layer graphene devices. The observation of CNP in negative gate voltage (n-type conduction) suggests the excess of electron (negative) charge carriers in graphene. The excess of n-type of carriers is certainly due to current annealing and possibly from the oxide layer. We do not expect any n-type doping during these experiments as all the measurements were performed in continuous high vacuum pumping ($<2\times10^{-6}$ Torr).

In order to gain quantitative insight into the current annealing effect on carrier mobility or residual carrier concentration, we used model developed by Kim et al.[16] The two terminal resistances ($R_{total}$) versus applied gate voltage can be fitted to the equation

$$R_{total} = R_{contact} + \frac{L}{We\mu \sqrt{n_0^2 + n^2}} \qquad (1)$$

Where e is electronic charge, L and W channel length and width, respectively. L =W (2 µm) in present case. The modulated carrier concentration (n) is given by n = α(Vg-$V_{Dirac}$), where α = 7.2x10$^{10}$ cm$^{-2}$. $V_{Dirac}$ is the gate voltage at which device resistance has highest value in R-(Vg) characteristics. Fitting of experimental data using equation 1, we can extract field effect mobility (µ), contact resistance ($R_{contact}$) and residual carrier concentration ($n_o$) or defect density in the oxide layer. Figure 3 (a) shows the fitting of resistance versus Vg (symbols) of the last three spectra (A, B, and C) in figure 1 (a). Fitting revealed contact resistance ($R_{contact}$) =1.18 kΩ, field effect mobility for spectra A, B, and C are 4375 cm$^2$/V.s, 4654 cm$^2$/V.s, and 4760 cm$^2$/V.s, whereas residual carrier concentration (defect density) 1.0x10$^{12}$ cm$^{-2}$, 5.9x10$^{11}$ cm$^{-2}$, and 5.5x10$^{11}$ cm$^{-2}$, respectively. Figure 3 (b) shows the fitting for again annealed spectra in figure 2. It is noticed that again current annealing reduced the mobility to 4059 cm$^2$/V.s from 4800 cm$^2$/V.s and $n_o$ increased 6.6x10$^{11}$ cm$^{-2}$, from 5.3x10$^{11}$ cm$^{-2}$. Further pumping of 2 to 16 hours reduced the defects concentration from 6.1x10$^{11}$ cm$^{-2}$ to 5.7x10$^{11}$ cm$^{-2}$, and improves the carrier mobility from 4404 cm$^2$/V.s to 4648 cm$^2$/V.s, respectively. These analysis show that just after current annealing we have higher defect density concentration and low mobility. The decrease in mobility with increase in carrier concentration after annealing is expected due to higher scattering of charge carriers and perturbation/disorder caused by beneath oxide layer. It is also seen that the fitting of experimental data for just after annealing is well fitted whereas later far from the CNP its deviates from experimental data. Independent fitting for hole and electron side of curve C in figure 3 (a) results higher hole mobility (4100 cm$^2$/V.s) comparative to electron mobility (3500 cm$^2$/V.s). Such asymmetry is understood due to imbalanced electron-hole injection from the graphene electrodes caused by the doping induced neutrality misalignment.[15,17] Deviation is higher in electron side, suggesting p-type doping[12,15,17,18] and support the CNP in positive gate side for these curves (C-in figure 1(a) and curve A in figure 3 A). In sample B and C, the field effect mobility values of 580 cm$^2$/V.s and 2250 cm$^2$/V.s have been achieved after current annealing, respectively.

The n-type conduction in current annealed graphene is understood in terms of negative charge carriers (electron) doping in graphene channels. High current induced charge trapping and de-trapping with time spend in also observed in metal oxide semiconductor (MOS) transistors in the past.[19,20] It is suggested that in annealed and/or stressed $SiO_2$ ($O_3 \equiv$Si-O-Si$\equiv O_3$) layers the trapping sites are cause by oxygen deficiency defects as a dangling silicon bond, $O_3 \equiv$Si·, and $O_3 \equiv$Si· ·Si$\equiv O_3$.[21] In present experiment, high current density (>3x10$^8$A/cm$^2$) passes through in graphene channel for longer time results high temperature annealing of graphene and defect creation in oxide layer which are responsible in charge trapping in oxide. The defect $O_3 \equiv$Si· ·Si$\equiv O_3$ proposed to undergo an internal electron transfer resulting in a dipole-like centre. The observed higher hysteresis in current annealed graphene (figure 2) may come due to formation of such dipole centres in annealed devices. The observation of n-type doping is expected due to charge trapping in defects formed in oxide layers and recovery/de-trapping with time results in the shift of CNP towards positive voltage. Recently, highly n-type doping behaviour also observed in high temperature (200$^0$C) annealed graphene supported on SiO$_2$/Si surface for 20 hours in vacuum.[22] The results are interpreted with the help of electronic energy calculations suggesting the electronic charge transfer from modified SiO$_2$ layer to graphene. These results support our interpretation doping of graphene channel from the excess charge in modified oxide layers. A careful look to the reported results on current annealed graphene[7,11] shows the CNP in negative gate voltage but origin was not discussed and possibly due to change in oxide layer. Current annealing is an efficient method to remove unintentionally impurities but we have to understand other effects in graphene and oxide to develop graphene device processing method. We believe that our results will be very useful to understand the mechanism of current annealing effects and in development of graphene based field effect transistors/devices.

In summary, we reported the observation of n-type conductivity in current annealed graphene and few layer graphene. With time spend, samples start to recover p-type conduction and saturate to a fixed gate voltage. N-type conductivity after annealing is due to the charge trapping in oxide layer during current annealing and shift in CNP in positive side due to de-trapping (reduction) of charge carriers.

We gratefully acknowledge the financial support from the Department of Science and Technology (DST), New Delhi and Agence Nationale de la Recherche (ANR), France. We thank Prof. Walter Escoffier, LNCMI, Toulouse, France for enlightening discussions and his group members for providing the measurements facilities.

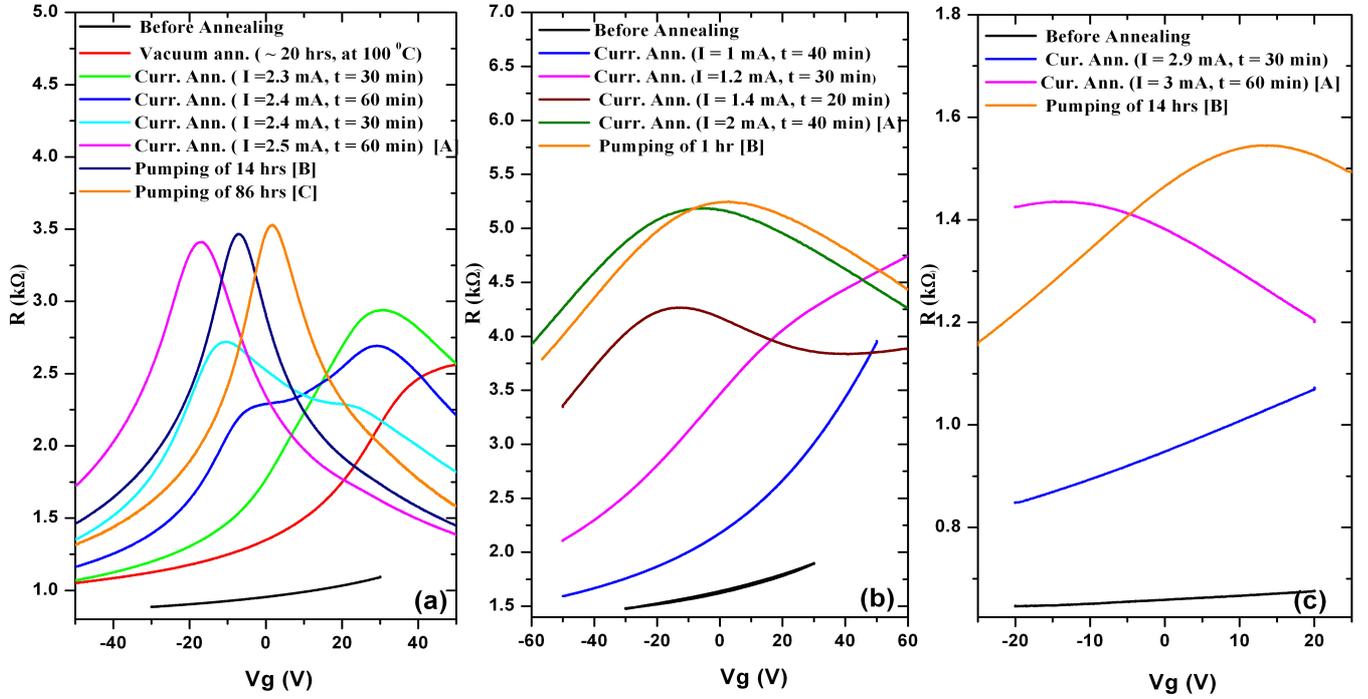

**FIG. 1.** (Colour online) Graphene channel resistance change with the sweep of applied gate voltage measurement for before annealing and after annealing at indicated current/time for **g**raphene devices, called as sample A (mono-layer graphene), sample B (~5-6 graphene layers) and sample C (>6 graphene layers)

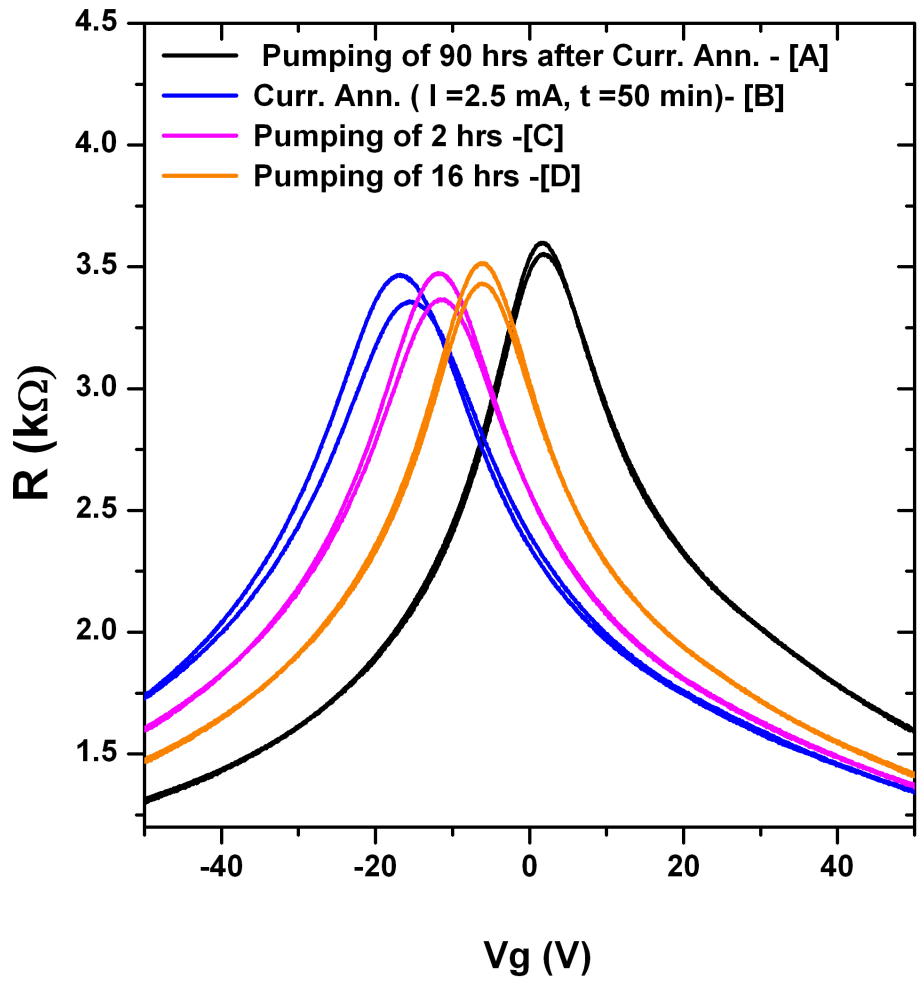

**FIG 2.** (Colour online) Resistance versus applied gate voltage for again current annealing of sample A.

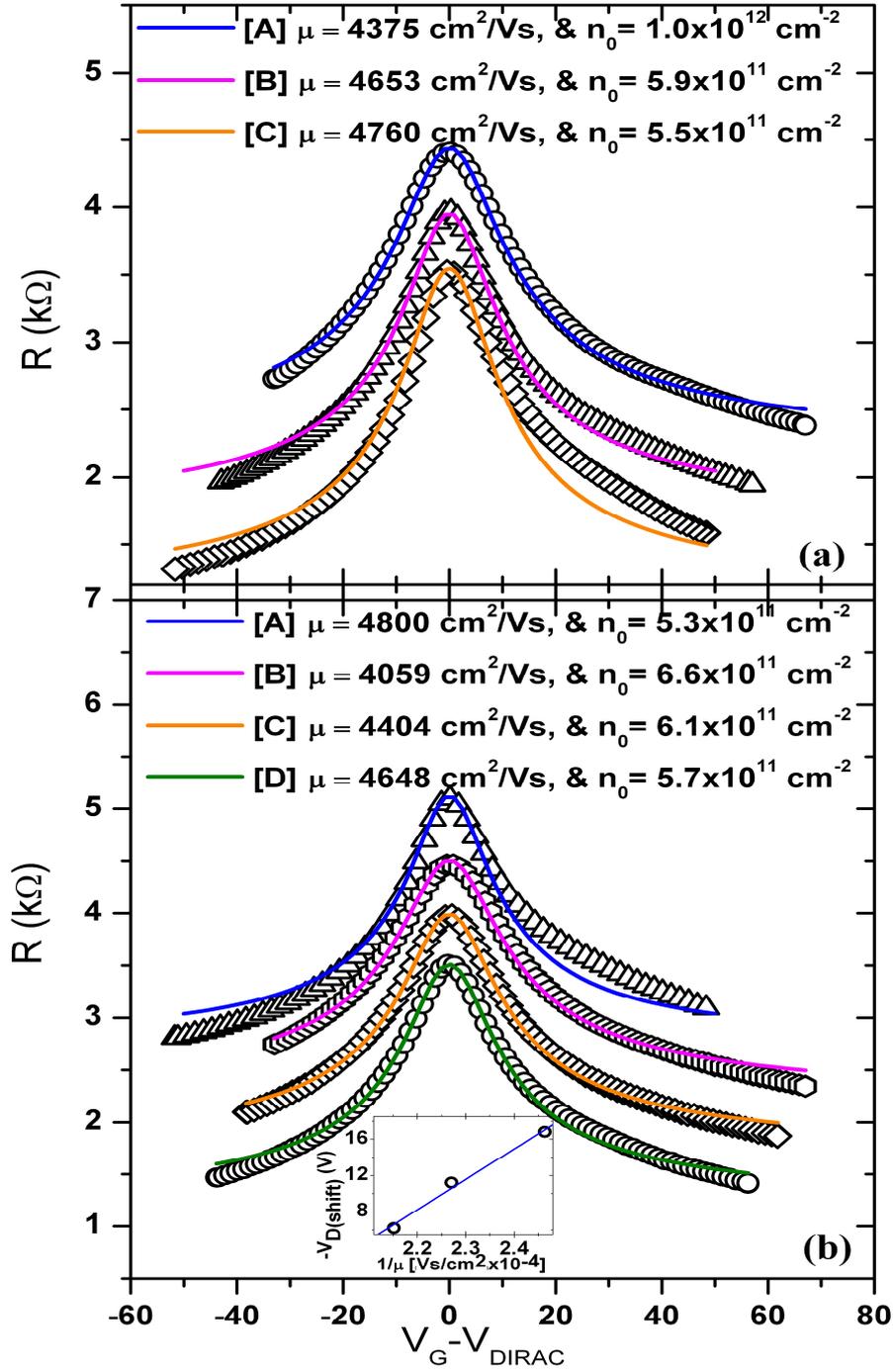

**FIG. 3.** (Colour online) $R_{total}$ versus $V_G-V_{DIRAC}$ experimental data (symbols) and fitting (solid lines). Spectra have been shifted (0.5kΩ) with respect to other for better clarity. Extracted field effect mobility (μ) and residual carrier concentration ($n_o$) or defect density are mentioned after each treatments.

13